\documentclass[pre,twocolumn,aps,superscriptaddress]{revtex4}
\usepackage[version=3]{mhchem} 
\usepackage[T1]{fontenc}       
\usepackage{epsfig,amsmath,amssymb,graphicx,color,calc,epstopdf}
\usepackage{color}
\usepackage{inputenc}

\newcommand{\br}{\mathbf{r}}
\newcommand{\ud}{\mathrm{d}}

\newcommand{\beq}{\begin{equation}}
\newcommand{\eeq}{\end{equation}}
\newcommand{\ba}{\begin{align}}
\newcommand{\ea}{\end{align}}

\makeatletter 
\@addtoreset{equation}{section}
\makeatother  

\newcommand{\zy}[1]{#1}

\begin{document}

\title{Recent Advances in the Theory and Simulation of Model Colloidal Microphase Formers}
\author{Yuan Zhuang}
\affiliation{Department of Chemistry, Duke University, Durham,
    North Carolina 27708, USA}
\author{Patrick Charbonneau}
\affiliation{Department of Chemistry, Duke University, Durham,
    North Carolina 27708, USA}
\affiliation{Department of Physics, Duke University, Durham,
    North Carolina 27708, USA}
\email{patrick.charbonneau@duke.edu}

\begin{abstract}
    This mini-review synthesizes our understanding of the equilibrium behavior of particle models with short-range attractive and long-range repulsive (SALR) interactions. These models, which can form stable periodic microphases, aim to reproduce the essence of colloidal suspensions with competing interactions. Ordered structures, however, have yet to be obtained in experiments. In order to better understand the hurdles to periodic microphase assembly, marked theoretical and simulation advances have been made over the last few years. Here, we present recent progress in the study of microphases in models with SALR interactions using liquid-state theory and density-functional theory as well as numerical simulations. Combining these various approaches provides a description of periodic microphases, and give insights into the rich phenomenology of the surrounding disordered regime. Three additional ongoing research directions in the thermodynamics of models with SALR interactions are also presented.
\end{abstract}

\maketitle

\section{Introduction}
\begin{figure*}
    \includegraphics[width=0.8\textwidth]{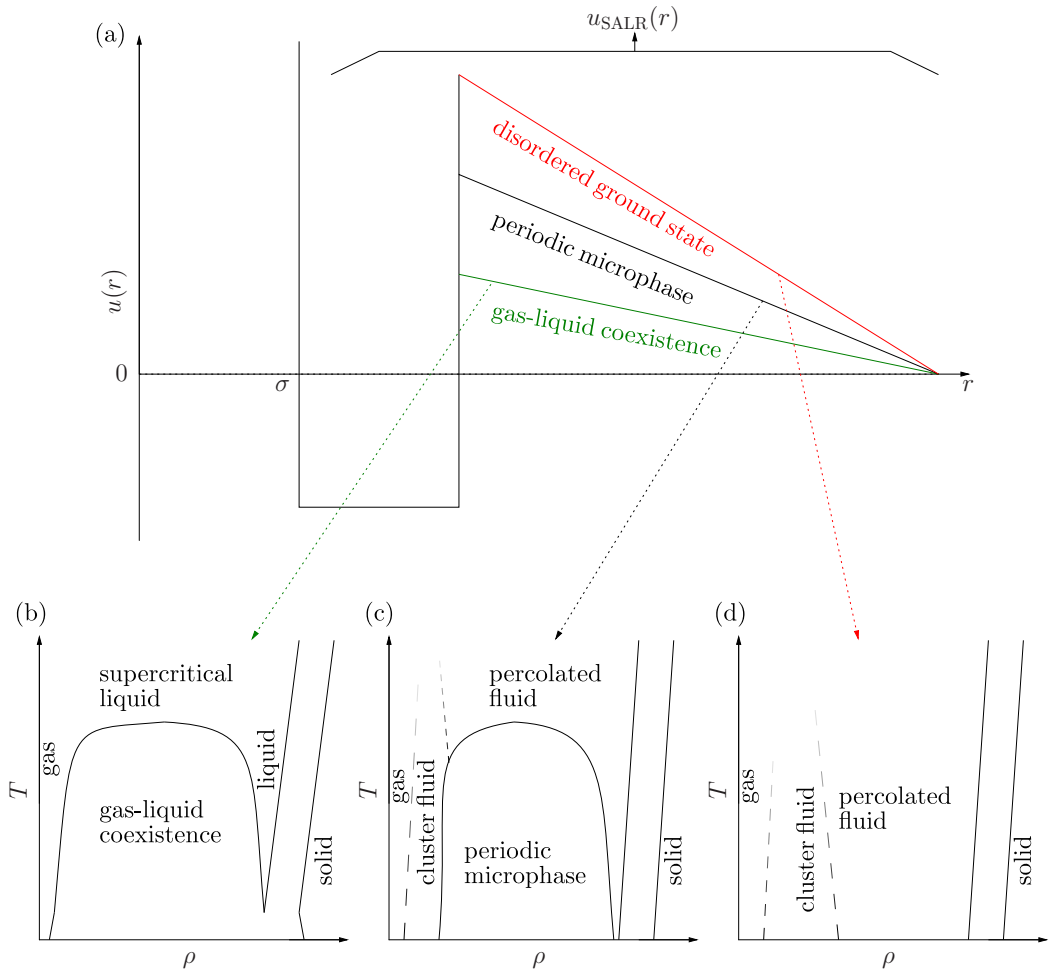}
    \caption{(a) Schematic SALR pair interaction potential, $u(r)$, as a function of the inter-particle distance, $r$. This particular potential has a hard spherical core of diameter $\sigma$ and a SALR contribution, $u_\mathrm{SALR}$, comprised of a square-well attraction and a linear repulsive ramp of different amplitudes. The resulting phase behavior depends on the repulsion strength.(b) For a weak repulsion the system is essentially a simple liquid; it displays a standard gas--liquid coexistence regime. (c) For a sufficiently strong repulsion a periodic microphase regime emerges and is surrounded by a complex \zy{disordered} regime \zy{that includes cluster and percolated fluids}. (d) For a very strong repulsion periodic microphases are thought to be absent even at low temperatures, leaving only \zy{a complex disordered regime at low densities}.  Solid lines indicate first order transitions and other lines denote crossovers. }
    \label{fig:potential}
\end{figure*}
\begin{figure}
    \includegraphics[width=0.5\textwidth]{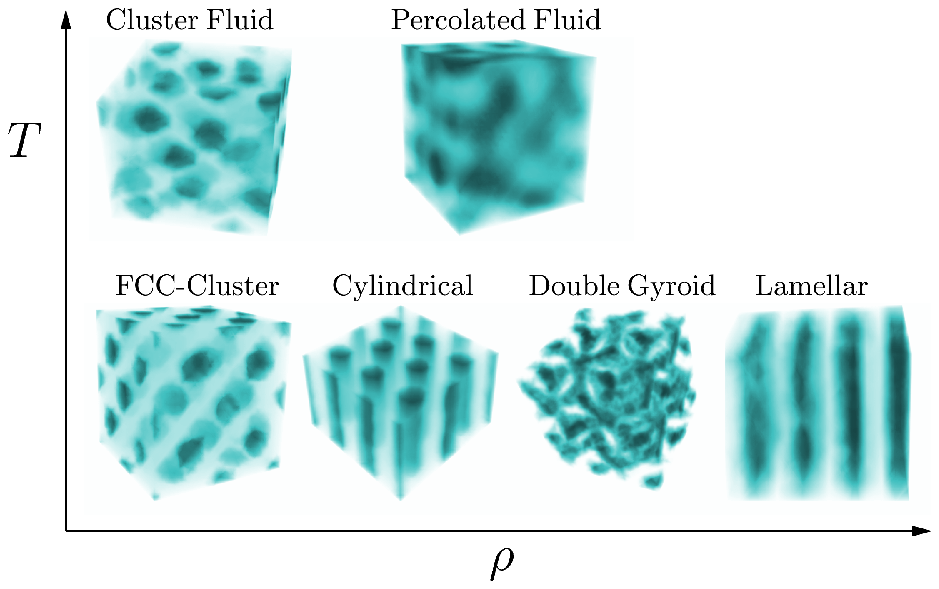}
    \caption{Coarse-grained density profiles of \zy{some of the} typical phases found in models with SALR interactions. The temperature $T$ and  density $\rho$ axis give a rough estimate of the phase positions; see Fig.~\ref{fig:potential} for more details. At high temperatures the system is a homonegenous fluid \zy{(not shown)}; at intermediate temperatures clusters first form and then percolate \zy{as density increases}; at low temperatures the system forms periodic microphases of various morphologies.}
    \label{fig:phases}
\end{figure}

Microphases are thermodynamically stable mesoscale structures that result from competing short-range attractive and long-range repulsive (SALR) inter-particle interactions. They universally replace gas-liquid (or equivalent) coexistence in systems with attraction alone, irrespective of the \zy{microscopic forces that give rise to the SALR interaction}\cite{Brazovskii1975,Seul1995,Ciach2013} (see Fig.~\ref{fig:potential}). As a result, microphases are observed in systems as diverse as magnetic alloys\cite{Seul1995}, Langmuir films\cite{Keller1986}, and protein solutions\cite{Stradner2004}. When microphases with long-range, periodic order develop, the resulting structures are both elegant and useful (see Fig.~\ref{fig:phases}). Block copolymers\cite{Bates1990,Bates1999,Kim2010}, for instance, can form a rich array of periodic morphologies, such as clusters, lamellae, and gyroid\cite{Leibler1980,Matsen1996}, with industrial applications in drug delivery\cite{Kataoka2001,Rosler2001}, nanoscale patterning\cite{Li2006,Krishnamoorthy2006}, and lithography\cite{Hawker2005,Tang2008}, \zy{among others}.

The earliest attempts at understanding microphase formation \zy{were driven by diblock copolymer experiments}. The resulting phase diagram, which includes periodic cluster crystal, cylindrical, gyroid and lamellar phases\cite{Leibler1980}, was further enriched hand in hand with experimental advances in those systems.\cite{Bates1990,Bates1999,Kim2010,Matsen1996} 

With increasing control over inter-particle interactions in colloidal suspensions and inspired by the apparent universality of microphase formation in systems with SALR interactions, various attempts at obtaining periodic structures in these setups have since also been made\cite{Stradner2004,Jordan2014,Campbell2005,Klix2010,Zhang2012b,Klix2013}. A mix of experimental challenges and theoretical limitations, however, have so far stymied obtaining periodic structures. First, colloidal interactions are harder to control than initially thought\cite{Campbell2005,Klix2010,Klix2013}. It was first assumed that depletion attraction and screened charge repulsion between colloids would suffice, but correlations in charge redistribution\cite{Klix2013} and other non-idealities play a significant role. Second, even if perfect experimental control were to be obtained, a fine conceptual grasp of particle microphase assembly seems necessary for it to be achieved. Indeed, even numerical simulations struggle to order microphases, and whether the difficulty is due to equilibrium\cite{Toledano2009,Liu2005} or out-of-equilibrium dynamics\cite{deCandia2006,Tarzia2007,Charbonneau2007,Schmalian2000,Geissler2004}, or even defect annealing \cite{Zhang2006,deCandia2011,DelGado2010} has been difficult to resolve. 

To make progress on this fundamental and applied materials question, a more careful determination of the equilibrium phase behavior of colloidal models with SALR interactions is thus essential. Only with this information in hand, can one hope to fully resolve the assembly dynamics of these models and to properly guide experiments. Fortunately, sizable advances have recently been accomplished using two main analytical approaches--liquid-state theory and density-functional theory--as well as numerical simulations based on novel methodological approaches. 
In this mini-review, we present the modalities and capabilities of these advances as well as the insights they provide into the equilibrium behavior of particle-based SALR models. We also present three related questions that remain open.

\section{Liquid-state theory}
\begin{figure}
    \includegraphics[width=0.5\textwidth]{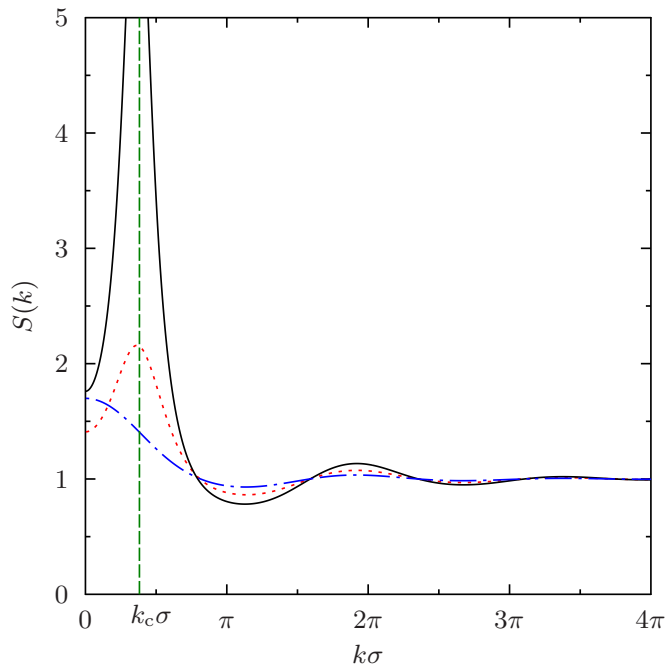}
    \caption{Schematic low-temperature structure factors for a model with SALR interactions for  different densities. At low density, the system is a homogeneous fluid of particles (blue, dot-dashed line); increasing density results in clustering, as seen by the emergence of a low-wavevector peak at $k_\mathrm{c}$ (red, dashed line); after entering the periodic microphase regime, i.e., beyond the $\lambda$ line, $S(k)$ diverges at $k_\mathrm{c}$ (black, solid).}
    \label{fig:sk}
\end{figure}

The structure factor, $S(k)$, of a homogeneous fluid is finite for all wavevector $k$ \zy{(Fig.~\ref{fig:sk})}. The appearance of a divergence in $S(k)$ thus signals an instability to density fluctuations. For a $k=0$ divergence macroscopic phase separation ensues, which in liquid-state theory typically identifies the gas-liquid spinodal.~\cite{Hansen1986} A $k>0$ divergence, by contrast, signals instability with respect to mesoscale fluctuations, which indicates the presence of a periodic microphase regime \zy{(See Fig.~\ref{fig:sk})}. In this section we present different ways in which classical liquid-state ideas have been used to understand the properties of models with SALR interactions.

\begin{itemize}
	\item 
	\textbf{Structure factor and periodic microphases:} A sign of the existence of periodic microphases is the presence of a diverging peak at a wavevector $k_\mathrm{c}\in(0,2\pi/\sigma)$ of $S(k)=\frac{1}{N}\left< \sum_{ij}e^{-i \mathbf{k}\cdot \br_{ij}}\right>$, where $k=|\mathbf{k}|$ and \zy{$\sigma$ is the particle diameter}. A $k_\mathrm{c}$ divergence \zy{captures} the long-range ordering that takes place when the repulsion \zy{of the SALR interaction} becomes sufficiently strong.~\cite{Sciortino2004}  The Ornstein--Zernike equation relates $S(k)$ to the direct correlation function as $S(k)=\frac{1}{1-\rho \hat{c}(k)}$, where $c(k)$ is a quantity commonly approximated in liquid-state theory.
\end{itemize}

\subsection{Zero-temperature approximation}
The simplest way to infer the presence of a low--$k$ divergence in $S(k)$ for a given model is from \zy{analyzing} its low-temperature behavior~\cite{Ciach2013}. The problem then reduces to determining whether the lowest-energy structure at a given density, i.e., the energetic ground state, is modulated or not. For a microphase regime to exist, a sufficiently strong repulsion is needed to effectively compete with inter-particle attraction and thus macroscopic condensation~\cite{Ciach2013}. Rough and coarse-grained structural approximations provide an approximate diagnostic as to whether a system is a microphase former or not\cite{Ciach2013}. More sophisticated yet still coarse-grained treatments find that modulated ground states can be as as morphologically diverse as their higher-temperature counterparts, and include cluster crystals, cylinders and lamellae as well as their inverses.~\cite{Wu2006}

Using a fully microscopic description, de Candia et al.~have further shown that the ground state structure is not only determined by the choice of periodic morphology, but also by the commensurability of the system size with that morphology~\cite{deCandia2006}. The zero-temperature stability of a given \zy{symmetry} at fixed density thus strongly depends on its occupancy (or thickness) and periodicity. (We will see below that a finite-temperature counterpart of this effect plays a key role in simulations.) Interestingly, the same work showed that for some models the low-density ground state \zy{is made of} cylindrical Coxeter helices rather than clusters. The \zy{low-temperature regime} of some models may thus not display any clustering, in contrast to standard expectations~\cite{Bollinger2016a}. The genericness of low-density helices, however, remains undetermined.
 
Although they are useful indicators of microphase formation, periodic ground states do not suffice to describe the finite-temperature behavior of models with SALR interaction. First, increasing temperature changes the internal structure of the microphase features. Particles may indeed display local FCC (or other) order at zero temperature, but be liquid-like at finite temperatures. Second, equilibrium feature sizes and shapes change with temperature. Third, temperature further affects the relative stability of the various periodic microphase morphologies.

\subsection{\zy{Finite--temperature} estimates}
In order to explore microphase formation in models with SALR interactions at finite temperatures, one needs a more versatile estimate of $S(k)$. A standard approximation scheme for the direct correlation function is the random phase approximation~(RPA), which treats the SALR \zy{interaction} as a perturbation to the hard-core repulsion between particles~\cite{Hansen1986} (Fig.~\ref{fig:potential}). 
The resulting $\hat{c}^{\mathrm{RPA}}(k)$ estimates the divergence of the liquid $S(k)$ and hence the boundary of the periodic microphase regime, i.e., the $\lambda$-line. Such analysis reveals that the $\lambda$-line, like the critical point \zy{in the simple liquid regime}, gets depressed to lower temperatures when the relative contribution of the long-range repulsion increases with respect to the short-range attraction.~\cite{Bomont2012} 

RPA has also been used to estimate the onset of clustering, which lies outside the $\lambda$-line~\cite{Bomont2010,Kim2011,Bomont2012,Bomont2012b,Bollinger2016a,Bollinger2016b}, by considering the growth of a finite, low-$k$ peak in $S(k)$. This clustering is analogous to micelle formation in the sense that the average cluster size first grows quickly but continuously over a small density range, and that the cluster size then depends only relatively weakly on density. The typical cluster size and other features can thus be theoretically estimated.
 
Note that other \zy{standard} approximations, such as self-consistent integral equation~\cite{Bomont2012, Bomont2012b}, hybrid mean spherical approximation (HMSA)~\cite{Kim2011,Bomont2012b}, single phase entropy rule~\cite{Lee2010}, and hypernetted chain (HNC)~\cite{Broccio2006,Kim2011,Costa2011,Bomont2012b,Cigala2015}, have been used to estimate the direct correlation function, in complement to RPA. Of these only a pure HNC description does not predict a $\lambda$-line.\cite{Bomont2012} For the others, the results remain qualitatively similar \zy{to RPA}. Although numerical predictions depend slightly on the choice of approximation, no scheme is clearly quantitatively superior. 

\subsection{Strong Repulsion Regime}
For very strongly repulsive interactions, particle addition to even relatively small clusters is \zy{energetically} unfavorable. Because the resulting aggregates are purely repulsive, irregularly shaped and can display a \zy{wide} size dispersity\cite{Wu2004}, the system is then thought to remain disordered, even at low temperatures, for a broad range of densities. The existence of a Wigner glass-like behavior in this regime has further been proposed\cite{Broccio2006,Sciortino2004}. An estimate of the repulsion onset necessary for such a regime to emerge has been obtained from balancing the attractive and repulsive contributions\cite{Zhuang2016a}, but its theoretical consideration remains incomplete.

\section{Density functional theory}
Density functional theory~(DFT) expresses the \zy{Helmholtz} free energy, $F$, of a system as a functional of its density profile, which is then \zy{tuned to minimize $F$}. A simple DFT provides the spinodal instability of a model with SALR interactions in its homogeneous phase from the divergence of its compressibility.~\cite{Archer2007a} In order to capture the emergence of finite $k$ modulations in SALR models, however, richer DFT formulations are needed. In this section we describe different such schemes that have been used to describe the disordered and the \zy{periodic} microphase regimes.

\begin{itemize}
	\item \textbf{Density functional theory:} The free energy of a liquid can be subdivided between ideal and an excess contributions, i.e., $F^{\mathrm{DFT}}(\beta,\mathbf{\rho})=F_{\mathrm{id}}(\mathbf{\rho})+F_{\mathrm{ex}}(\mathbf{\rho})$ for a fluid at a number density $\rho$. A common approximation consists of taking the local liquid structure to be that of a hard-sphere (HS) fluid at the same density with a perturbative correction, i.e., $F_{\mathrm{ex}}(\mathbf{\rho})=F_{\mathrm{ex},\mathrm{HS}}(\mathbf{\rho})+\iint\ud\br\ud \br' \mathbf{\rho}(\br)\mathbf{\rho}(\br') u_{\mathrm{SALR}}(|\br-\br'|)$.
\end{itemize}

\subsection{Disordered Mesophases}
The simplest mesoscale structures formed by models with SALR interactions is the clustering of the low-density gas. A DFT strategy for studying this effect assumes a uniform distribution of clusters and isolated particles, which is akin to treating the system as a low-density binary fluid mixture~\cite{Sweatman2014}. Assuming the excess free energy to be that of a hard-sphere \zy{binary system}, clustering is deemed to emerge when the mixture free energy is lower than that of a uniform liquid of isolated particles. The resulting analysis describes a steep but continuous transition into the cluster fluid regime upon increasing density. This micellization-like transition is similar to that observed in \zy{$S(k)$}-based appproaches. 

DFT can also \zy{give insights into the relative} stability of \zy{various} cluster shapes.~\cite{Wu2005} Assuming that the main free energy of the clusters can be split between surface and core contributions, different cluster \zy{morphologies} can be compared. This approach shows that models with SALR interactions can result in cluster shapes that are far from spherical.

\subsection{Periodic Mesophases}
Two main DFT approaches have been used to study periodic microphases.

The first DFT scheme analyzes a family of models after mapping their microscopic Hamiltonian to a density field akin to the Landau--Brazovskii free energy functional~\cite{Ciach2008,Ciach2010,Pekalski2014}. The approximate mapping between the microscopic description and the field theory is achieved by encoding the interaction potential using an estimate of the correlation between two density fields. The resulting free energy expression is then minimized with respect to a density profile. This last operation that can be simplified by considering the symmetry of the various mesoscale morphologies.  At low temperatures, periodic microphases, including cluster crystal, cylindrical, double gyroid and lamellar phases, are found to be more stable than the homogeneous fluid. Interestingly, comparing the double and simple gyroid phases, which both are bicontinuous phases with Ia3d symmetry, reveals that only the former is thermodynamically stable in these systems~\cite{Ciach2008}. 

The second DFT scheme grids space into cubes that are parameterized by a density value. The resulting free energy expression is then numerically minimized with respect to the density profile. Because this approach is very sensitive  to the initial input profile and system size, however, both of these aspects must be treated carefully.  This challenge is reminiscent of the microphase occupancy difficulty encountered in the ground state determination. Although only the phase behavior of two-dimensional models has been determined thus far~\cite{Archer2008,Chacko2015}, nothing fundamentally prevents its application to three-dimensional systems. 

\section{Simulation Approaches}
\begin{figure*}
    \includegraphics[width=\textwidth]{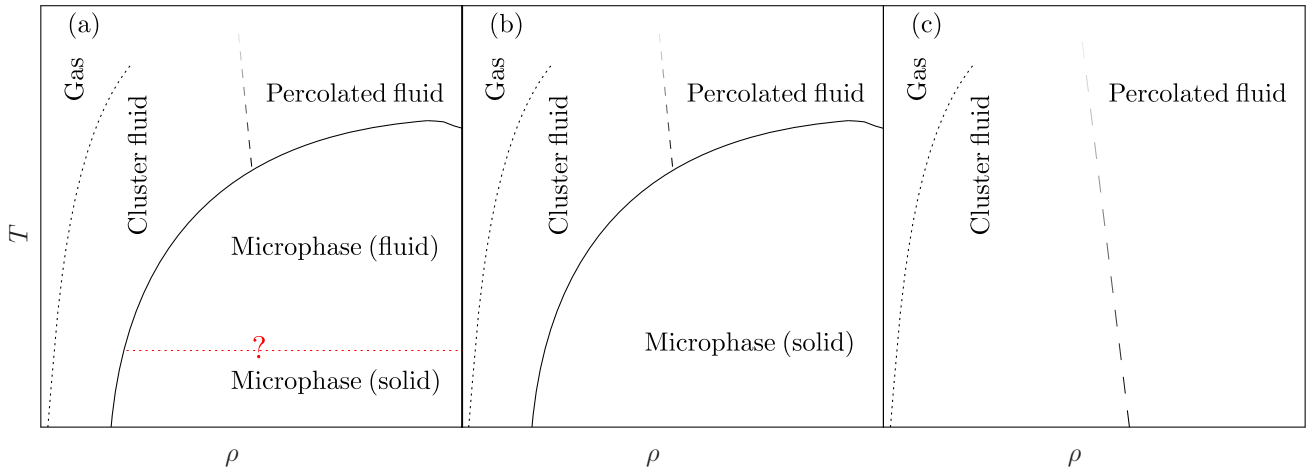}
    \caption{Schematic temperature-density, $T$--$\rho$, phase diagrams for models with SALR interactions. \zy{Solid lines indicate first order transitions and other lines indicate crossovers.} (a) For relatively wide attraction ranges, periodic microphases display fluid-like local order near their melting, but lowering temperature likely gives rise to solid-like ordering. (b) For short attraction ranges, the entire periodic microphase regime is expected to display local solid-like order. (c) For very strong repulsion ranges, no periodic microphases are expected.}
    \label{fig:phase_diagram}
\end{figure*}
Simulating \zy{models with SALR interactions} is particularly challenging because dynamically sluggish, disordered regimes and large finite-size effects interfere with equilibration. Only recently have methods to surmount some of these challenges become available. In this section, we consider how disordered and periodic microphases have been studied in simulations.

\subsection{Disordered Mesophases}
Theoretical analysis of clustering in models with SALR interactions are complemented by simulations~\cite{Broccio2006,Imperio2006,Bomont2010,Kim2011,Bomont2012b,Cigala2015,Zhuang2016,Bollinger2016a}. These studies have confirmed that clusters are more stable than the homogeneous liquid of monomers over a broad range of densities and temperatures. Like micelle formation, clustering develops sharply but continuously; results that suggest otherwise likely suffer from poor sampling.

Unlike micelles\cite{Bug1987}, however, clusters from models with SALR interactions have a size distribution that is fairly wide and distinct from Gaussian~\cite{Sciortino2004,Sciortino2005,Godfrin2013,Mani2014,Mani2015,Ruzicka2015,Ruzicka2015b,Zhuang2016}.  The study of diblock copolymers suggests that such clusters can be subdivided into various types.~\cite{Bug1987,Nelson1997,Desplat1996} 
Because similar distributions are observed in \zy{particle-based} microphase formers~\cite{Godfrin2013,Godfrin2014,Ruzicka2015,Ruzicka2015b,Zhuang2016}, clustering then also likely competes with elongation and crystallization. Particle models indeed display a remarkably rich behavior beyond clustering, with the clusters first elongating and then percolating while \emph{still outside} of the periodic microphase regime. 

The complexity of the clustering regime might affect the formation of periodic microphases. For instance, the assembly of cluster crystals was first proposed to result from the crystallization of \zy{a fluid of} roughly spherical and monodisperse clusters\cite{Sciortino2004}, but simulations indicate that the wide cluster-size distribution necessitates important rearrangements to the clusters for crystallization to \zy{proceed}\cite{Zhuang2016}. Because such effects can be difficult to capture analytically, simulations have been (and will \zy{likely} remain) instrumental in elucidating this assembly process. Fortunately, the clustering phenomenology seems to be fairly robust to the choice of SALR interactions and sampling dynamics, be it Brownian~\cite{Sciortino2005,Toledano2009,Mani2015}, MD~\cite{Sciortino2004,Mani2014} or MC ~\cite{Toledano2009,Godfrin2013,Godfrin2014,Ruzicka2015,Ruzicka2015b,Zhuang2016}, a certain physical universality is thus to be expected.

Cluster percolation in simulations~\cite{Sciortino2004,Valadez2013,Zhuang2016} is found to shift to smaller densities when the long-range repulsion increases~\cite{Valadez2013}. Although percolation is not a thermodynamic phase transition, it nonetheless gives rise to a marked change in the system's structural relaxation\cite{Zhuang2016} and is especially important for understanding its physical properties, such as conductivity and rheology. The equivalent phenomenon in microphase-forming diblock copolymers has found uses in batteries\cite{Croce1998}, fuel cells\cite{Steele2001}, and catalysis\cite{Lu2004}, \zy{but it remains to be similarly exploited in colloidal suspensions.} Note that percolation need not, however, \zy{necessarily} lead to gel formation, because system-wide rearrangements can still persist on microscopic timescales\cite{Mossa2004,Sciortino2005,Zhuang2016}.
Yet although \zy{percolation} might \zy{contribute to} the dynamical challenge of assembling periodic microphases, few studies has yet focused on this effect. 

As the repulsive contribution of a SALR interaction increases so does cluster size heterogeneity~\cite{Godfrin2014}. This effect likely plays a role in the system remaining disordered at low temperatures~\cite{Sciortino2004,Toledano2009}. Despite various models having been simulated\cite{Godfrin2014,Toledano2009,Zhuang2016}, more studies are needed to fully understand this regime (see Fig.~\ref{fig:phase_diagram} for \zy{a schematic phase diagram}).

\subsection{Periodic Microphases}
From simulations, it has been observed that two qualitatively different types of periodic microphases exist (see Fig.~\ref{fig:phase_diagram} for \zy{schematic phase diagrams}). On the one hand, for systems with wide attraction ranges, periodic microphases display local fluid-like ordering near the microphase melting regime~\cite{Zhuang2016}. At lower temperatures, however, local ordering likely develops. On the other hand, for systems with very short range attraction ranges, periodic microphases display local crystal-like order over their entire periodic microphase regime\cite{deCandia2006}. The distinction \zy{between the two types} likely echoes the gas-crystal coexistence line in systems with purely attractive interactions.

Studying the periodic microphase regime in further details suffers from two main difficulties: (i) various morphologies ought to be considered, and, more importantly, (ii) minimizing the free energy requires relaxing the thickness and periodicity these morphologies. This problem is similar to that of determining the energetic ground state\cite{deCandia2006,Pekalski2014,Zhuang2016}. Equilibrating periodic microphases thus faces a similar \zy{difficulty} as determining the equilibrium of vacancy of a standard crystal~\cite{Swope1992} and the occupancy of multiple-occupancy crystals\cite{Mladek2007,Mladek2008,Zhang2010b}. 

\begin{itemize}
	\item \textbf{Expanded thermodynamics:} \zy{Obtaining the free energy of} periodic microphases can be done by including an additional pair of conjugated variables: the lattice occupancy $n_{\mathrm{c}}$ and a chemical potential-like quantity $\mu_{\mathrm{c}}$. The differential form of the Helmholtz free energy, for instance, is then written as
	$	\ud F_{\mathrm{c}} = -S\ud T-P \ud V+\mu \ud N + \mu_{\mathrm{c}}N\ud n_{\mathrm{c}}$,
	with entropy $S$, temperature $T$, pressure $P$, volume $V$, chemical potential $\mu$ and number of particle $N$. At equilibrium we must recover $F(N,V,T)=F_{\mathrm{c}}(N,V,T,n_{\mathrm{c}}^{\mathrm{eq}})$, hence $F_{\mathrm{c}}$ must be minimized with respect to $n_{\mathrm{c}}$ to obtain $F$.
\end{itemize}

Early free energy simulations of the periodic regime of \zy{models with SALR interactions} used thermodynamic integration~(TI) from low-temperature ground states in order to extract finite-temperature information.\cite{deCandia2006} However, as mentioned above, temperature affects the local order as well as the thickness and periodicity of microphases. These features in turn affect the stability of different \zy{microphase morphologies}. These TI results are thus expected to suffer from relatively large finite-size effects and other difficulties. For instance, because the specific model considered in Ref.~\onlinecite{deCandia2006} lacks a stable cluster crystal energetic ground state, finite-temperature cluster crystals cannot be separately considered by \zy{this} TI scheme. The work nonetheless hints at the existence of an interesting interplay between disordered and periodic microphases. 

Recent methodological advances for studying periodic microphases entail using a specifically designed TI scheme for calculating the free energy of a given morphology and then explicitly minimizing the lattice occupancy.\cite{Zhuang2016} This TI method first connects an ideal gas under a (properly chosen) periodic external field to HS under the same field, and then to the full \zy{full} SALR interaction. Using this scheme, the stability and relative positions of the cluster crystal, cylindrical, double gyroid and lamellar phases \zy{have been determined for a couple of simple models}~\cite{Zhuang2016,Zhuang2016a}. 

\zy{Note that} the MC-based ghost particle switching developed by Wilding and Sollich~\cite{Wilding2013} has been shown to more efficiently sample and multiple-occupancy crystals than TI approaches. Because it allows the direct fluctuation and equilibration of the lattice occupancy, adapting ghost particle switching to models with SALR interactions will thus likely further improve computational capabilities.

\section{Open Questions and Future Directions}
Despite the recent sizable theoretical and computational advances in analyzing and characterizing the equilibrium properties of models with SALR interactions, important challenges remain to be addressed. Surmounting them will require further advances as well as a richer conceptual understanding of these systems. 

\subsection{High-density phase behavior}
\begin{figure}
    \includegraphics[width=0.5\textwidth]{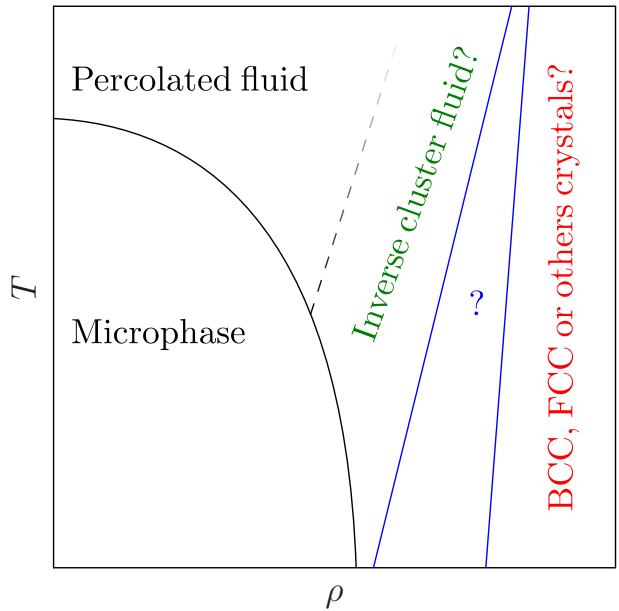}
    \caption{The high-density regime of models with SALR interactions is mostly uncharted.  Does an inverse cluster fluid exist? Does the microphase regime directly coexist with the crystal phase?  What is the symmetry of the various crystal phases? \zy{Solid lines indicate tentative first order transitions and other lines denote crossovers.}}
    \label{fig:high_density}
\end{figure}

As temperature increases, the behavior of models with SALR interactions must smoothly connect with that of \zy{models with only} core repulsion. For instance, a model with a hard core ought to \zy{behave like} simple hard spheres at high temperatures. Determining how does this behavior connects with the finite-temperature behavior of models with SALR interactions, however, remains difficult to construe (see Fig.~\ref{fig:high_density} for a proposal). One might expect the percolated liquid to eventually give rise to a phase with inverted clusters, the two regimes being separated by a void percolation transition. But how does the resulting disordered phase eventually give rise to a crystal? Relatively little is known or has been proposed for this regime from either theory or simulation. To the best of our knowledge, the only numerical attempt at extracting such information used TI from the hard-sphere, infinite-temperature reference~\cite{Mani2014}, but proper sampling for such a scheme is challenging. 

\subsection{Theoretical descriptions beyond mean-field}
Both analytical treatments presented in this mini-review are mean-field--like because they neglect contributions from fluctuations beyond mesoscale density modulations. Studies of diblock copolymers suggest that including fluctuations yields more accurate and informative phase information. Future theoretical efforts on SALR models will thus likely strive to reach beyond mean-field treatments. In two-dimensions, such corrections are especially important~\cite{Coto2012}, because periodic microphases with long-range order are not thermodynamically stable. Field-theory arguments dating back to Brazovskii further indicate that some corrections are also qualitatively important in three dimensions\cite{Brazovskii1975,Seul1995}. It is possible that a proper understanding of the very strong repulsion regime also requires descriptions that go beyond mean-field theory. 

\subsection{Anisotropic SALR particles}
Models with SALR anisotropic interactions are expected to exhibit an even richer phase behavior than their isotropic counterparts. Here again, inspiration comes from results in diblock copolymers. Rod-coil copolymers, for instance, display a wide array of  mesoscale morphologies, likely resulting from the coupling of local liquid-crystal ordering with mesoscale patterns\zy{\cite{Olsen2008,Pryamitsyn2004}}. The periodic mesophases of rod-like particles with SALR interactions is expected to exhibit a similar complexity, but have yet to be considered. From a more theoretical viewpoint, the order--disorder transition might also be weakened in these systems.

\section{Conclusions}
Our structural understanding of models with SALR interactions has markedly improved over the last few years. By piecing together advances from theory and simulation, a clearer picture of the equilibrium behavior of these models is finally emerging. In order for these results to inform experimental attempts at controlling the formation of periodic microphases, however, relating structural thermodynamics to assembly dynamics also has to be mastered. The coming years should thus provide a clear answer as to whether periodic microphases should be achievable in well-controlled experiments, and under what circumstances. The full materials promises of microphases in colloidal suspensions will then finally be within reach.

\begin{acknowledgements}
We acknowledge support from the National Science Foundation Grant no. NSF DMR-1055586 and from the Materials Research Science and Engineering Centers~(DMR-1121107).
\end{acknowledgements}

\end{document}